%% file: ms.tex
\pdfoutput=1
\documentclass{sig-alternate}
\usepackage[margin=1in]{geometry}

\usepackage{booktabs} % For formal tables

\usepackage[ruled]{algorithm2e} % For algorithms

\SetAlFnt{\small}
\SetAlCapFnt{\small}
\SetAlCapNameFnt{\small}
\SetAlCapHSkip{0pt}
\IncMargin{-\parindent}

\usepackage{graphicx}

\usepackage{hyperref} 
\usepackage{url}

\begin{document}
\pagestyle{empty}
\title{Progressive Data Science: Potential and Challenges}

\numberofauthors{10} %  in this sample file, there are a *total*
\author{
	% 1st. author
	\alignauthor
	Cagatay Turkay\\
	\affaddr{City, University of London, UK}\\
	\email{Cagatay.Turkay@city.ac.uk}
	% 2nd. author
	\alignauthor
	Nicola Pezzotti\titlenote{Nicola Pezzotti is also affiliated with Phillips Research, Eindhoven, NL}\\
	\affaddr{TU Delft, NL}\\
	\email{N.Pezzotti@tudelft.nl}
	% 3rd. author
	\alignauthor Carsten Binnig\\
	\affaddr{TU Darmstadt,  DE}\\
	\email{carsten.binnig@cs.tu-darmstadt.de}
	\and  % use '\and' if you need 'another row' of author names
	%4th
	\alignauthor Hendrik Strobelt\\
	\affaddr{IBM Research AI, USA}\\
	\email{hendrik@strobelt.com}
	% 5th. author
	\alignauthor Barbara Hammer\\
	\affaddr{Bielefeld University, DE}\\
	\email{bhammer@techfak.uni-bielefeld.de}
	% 6th. author
	\alignauthor Daniel A. Keim\\
	\affaddr{University of Konstanz, DE}\\
	\email{keim@uni-konstanz.de}
	\and  % use '\and' if you need 'another row' of author names
	% 7th. author
	\alignauthor Jean-Daniel Fekete\\
	\affaddr{INRIA, FR}\\
	\email{Jean-Daniel.Fekete@inria.fr}
		% 8th. author
	\alignauthor Themis Palpanas\\
	\affaddr{Paris Descartes University, FR}\\
	\email{themis@mi.\\parisdescartes.fr}
		% 9th. author
	\alignauthor Yunhai Wang\\
	\affaddr{Shandong University, CN}\\
	\email{cloudseawang@gmail.com}
	\and  % use '\and' if you need 'another row' of author names
	% 10th. author
	\alignauthor Florin Rusu\\
	\affaddr{University of California Merced, USA }\\
	\email{frusu@ucmerced.edu}
}
% There's nothing stopping you putting the seventh, eighth, etc.
% author on the opening page (as the 'third row') but we ask,
% for aesthetic reasons that you place these 'additional authors'
% in the \additional authors block, viz.
%\additionalauthors{Additional authors: John Smith (The Th{\o}rv{\"a}ld Group,
%	email: {\texttt{jsmith@affiliation.org}}) and Julius P.~Kumquat
%	(The Kumquat Consortium, email: {\texttt{jpkumquat@consortium.net}}).}
% Just remember to make sure that the TOTAL number of authors
% is the number that will appear on the first page PLUS the
% number that will appear in the \additionalauthors section.

\maketitle

% Title portion. Note the short title for running heads

% \affiliation{%
% \email{beranger@inria.fr}
% \affiliation{%
% \email{aprna_patel@rguhs.ac.in}
% \affiliation{%
%   \country{China}
% }
% \email{chan0345@tsinghua.edu.cn}
% \author{Ting Yan}
% \affiliation{%
%   \institution{Eaton Innovation Center}
%   \city{Prague}
%   \country{Czech Republic}}
% \email{yanting02@gmail.com}
% \author{Tian He}
% \affiliation{%
%   \institution{University of Virginia}
%   \department{School of Engineering}
%   \city{Charlottesville}
%   \state{VA}
%   \postcode{22903}
% \affiliation{%
%   \institution{University of Minnesota}
% \email{tinghe@uva.edu}
% \author{Chengdu Huang}
% \author{John A. Stankovic}
% \author{Tarek F. Abdelzaher}
% \affiliation{%
%   \institution{University of Virginia}
%   \department{School of Engineering}
%   \city{Charlottesville}
%   \state{VA}
%   \postcode{22903}

%\maketitle

\begin{abstract}
Data science requires time-consuming iterative manual activities. In particular, activities such as data selection, preprocessing, transformation, and mining, highly depend on iterative trial-and-error processes that could be sped-up significantly by providing quick feedback on the impact of changes. The idea of progressive data science is to compute the results of changes in a progressive manner, returning a first approximation of results quickly and allow  iterative refinements until converging to a final result. Enabling the user to interact with the intermediate results allows an early detection of erroneous or suboptimal choices, the guided definition of modifications to the pipeline and their quick assessment. In this paper, we discuss the progressiveness challenges arising in different steps of the data science pipeline. We describe how changes in each step of the pipeline impact the subsequent steps and outline why progressive data science will help to make the process more effective. Computing progressive approximations of outcomes resulting from changes creates numerous research challenges, especially if the changes are made in the early steps of the pipeline. We discuss these challenges and outline first steps towards progressiveness, which, we argue, will ultimately help to significantly speed-up the overall data science process.  
\end{abstract}


\input{Sec1-Introduction.tex}

\input{Sec2-ProgresivenessChallenges.tex}

\input{Sec3-Promisingfirststeps.tex}
\input{Sec4-Discussion.tex}
\input{Sec5-Conclusion.tex}

\section{Acknowledgments}

The authors would like to thank Schloss Dagstuhl – Leibniz Center for Informatics for their support in the organisation of the ``Dagstuhl Seminar 18411 - Progressive Data Analysis and Visualization''.

% Bibliography
\bibliographystyle{acm}
\bibliography{ms}

\end{document}

%% file: Sec1-Introduction.tex
\section{Introduction}

Data science is an iterative multi-stage knowledge discovery (KDD) process in which analysts start working with raw, often non-cleaned collections of data sources to derive context-relevant knowledge through the observations made and the computational models built. The overall process involves several labor-intensive trial and error steps within the core activities of data selection, preprocessing, transformation, and mining. This iterative, trial-based nature of the process often means that analysts spend significant amount of time on each stage to move through the analysis pipeline---to give an example, the interviews with enterprise analysts by Kandel et al.~\cite{kandel2012enterprise} report that even preparatory data wrangling steps can easily take more than half of the analysts' time, keeping them off from the rather creative and insightful phases of data analysis. In this paper, we argue how progressive methods, where approximate but progressively improving results are provided to analysts in short time, can transform how the KDD process is currently conducted when progressiveness is introduced within each step of the pipeline, and we introduce \textit{Progressive Data Science} as a novel paradigm.

The underpinning idea of progressive methods is to provide analysts with approximate, yet informative, intermediate responses from a computational mechanism in short time. The analysts are then supported to interactively investigate these early results and empowered to choose to either discard the chosen conditions due to suboptimal early results, or wait for a full-quality result with the chosen conditions following promising first observations. An illustrative example is an unsupervised clustering process where an analyst is trying to find groups within millions of high-dimensional data observations---a computation that takes considerable amount of time even with an efficient algorithm. To further complicate this, analysts would usually like to investigate several different distance metrics that will give them distinguishable, well-defined groups---a task that can easily become intractable if a single clustering run takes a few hours, if not days. In the progressive setting that we envision, an approximate clustering of the observations is provided as quickly as possible, and, if the initial results fail to provide evidence that any useful structure is captured, that distance function could be discarded immediately, saving the analyst precious time -- by not waiting for the full result -- and making the time available to try the next alternative distance function. 

Furthermore, the progressive approaches we envision do not only help to speed up individual steps of the KDD process, but, more importantly, allow data scientists to quickly revisit previous decisions and immediately see their effects on other steps. For example, in a classical setup, data has to be cleaned (by replacing missing values, removing outliers, etc.) before a data mining algorithm or a machine learning model is applied. In a progressive data science pipeline, we envision that a data scientist can start working on the early steps such as cleaning the data (removing obvious problems) and then move forward to the later steps (e.g., apply the clustering algorithm) already on the partially cleaned data. By looking at the result of clustering the data, new data errors might become visible to the data scientist (e.g., certain types of outliers). Based on these observations, the data scientist could revisit and alter the data cleaning step to remove these outliers and immediately see the effects on the clustering algorithm.

These two examples already showcase the vision for an iterative, high-paced progressive data science process that we argue for. Early promising examples of progressive approaches have been recently introduced in the database~\cite{agarwal2013blinkdb,vartak2015s}, machine learning~\cite{losing2016knn,rusu2016progressive,Rusu:SGDOLA,pezzotti2017approximated}, and visualization communities~\cite{fekete2016progressive,turkay2017designing,jo2018panene}. 
This paper aims to present a unifying vision through a rethinking of the widely adopted and influential KDD pipeline~\cite{Kriegel2017}, and introduces Progressive Data Science as a novel knowledge discovery paradigm where progressiveness is inherent in every step of the process. It is important to highlight that we are not the first ones to argue for progressive approaches within data science practices -- coming from a machine learning model building perspective, Vartak et al.~\cite{Vartak2016} demonstrate how incremental and human-in-the-loop model building methods could transform model building with Xin et al.~\cite{Xin2018} discussing how such processes can be optimized, while Hellerstein et al.~\cite{hellerstein1999control} and Raman et al.~\cite{raman1999spreadsheets} discuss the role of interactive online processing from a databases and data analysis perspective. Our position here builds on these ideas and discusses a more comprehensive perspective that considers the whole KDD pipeline rather than emphasizing certain parts of it.

To present our position, we base the discussion on the individual stages of the KDD pipeline, from data selection, preprocessing, and transformation, to data mining and evaluation, and present how progressiveness can be introduced within each stage and discuss how changes in one stage impact subsequent stages in this progressive setting. In the remainder of this paper, we visit each stage, identify potential opportunities and challenges that lie ahead in integrating progressiveness, and discuss the benefits and implications of this transformation through examples. We then present a number of first promising steps in the database, machine learning, and visualization communities to open up further discussions in this high potential research area.

%% file: Sec2-ProgresivenessChallenges.tex
%\section{\texorpdfstring{Progressiveness: \\ Opportunities \& Challenges}}

\section[Progressiveness: Opportunities \& Challenges]{\texorpdfstring{Progressiveness: \\ Opportunities \& Challenges}{Progressiveness: Opportunities \& Challenges}}

\label{sec:challenges}
\begin{figure*}[t!]
\centering
\includegraphics[width=0.8\textwidth]{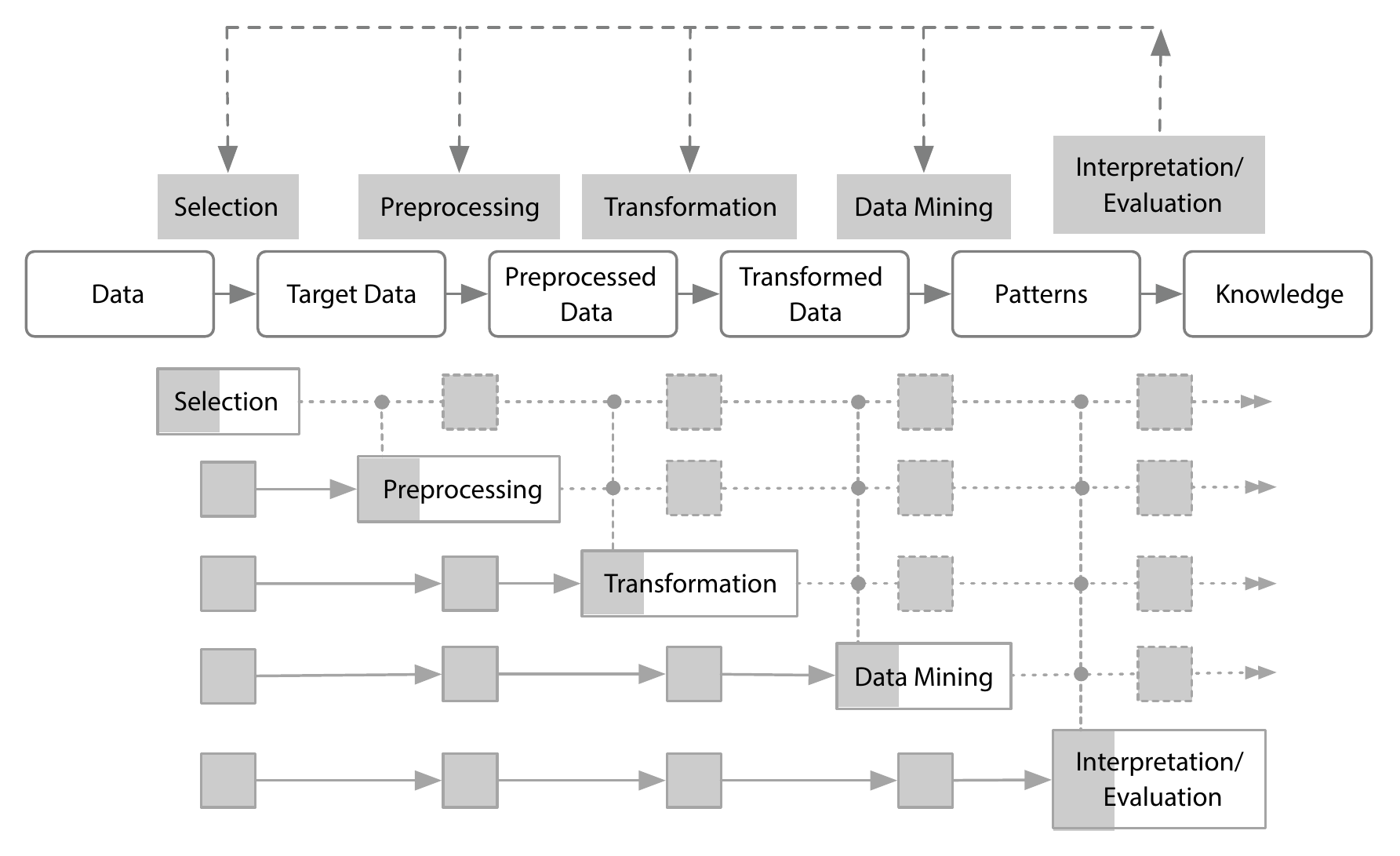}
\caption{We base our revised Progressive Data Science pipeline on the individual stages of the established KDD pipeline and present how progressiveness can be introduced within each stage. Notice that each stage operates on and produces data in a progressive manner, enabling analysts to effectively move upstream and downstream along the pipeline. Horizontally parallel lines indicate alternative runs, with approximate results at different stages (indicated by gray, dashed boxes) inform transitions between different KDD steps.}
\label{fig:kddPipeline}
\end{figure*}

For each stage in the KDD pipeline~\cite{Kriegel2017}, we identify opportunities for using progressive methods and present the implications of progressiveness on their input and output. Informed by and closely following the established KDD pipeline~\cite{Kriegel2017}, we are rethinking the whole process in a progressive manner, as illustrated in Figure~\ref{fig:kddPipeline}. In this paradigm, approximate results are produced (the dashed boxes indicate such intermediate results in Figure~\ref{fig:kddPipeline}) along multiple alternative runs that are ``tried'' progressively at different stages of the pipeline, as indicated by each horizontal line in Figure~\ref{fig:kddPipeline}. Each section includes concrete examples that provide a clear understanding of the involved challenges.

\noindent \textbf{A running example: } For the remainder of this paper, we will be using the same unsupervised clustering example that we touched upon in the introduction as a vehicle to talk through the concepts. In this example, let us consider a scenario where an analysts working for an online retailer who is trying to segment their customers according to different criteria. In this setting we are dealing with very large volumes of customer data that sit across disintegrated data sets where for each customer, the data consists of features with varying characteristics such as demographics, spending patterns, personal details to name a few. 

\subsection{Data Selection}
\label{sec:kdd_selection}
Data selection is typically the very first step of a KDD pipeline where users need to explore new data sets and decide whether or not a new data set is relevant for further investigation. In previous work, different approaches have been proposed to increase user efficiency during the data selection phase~\cite{lissandrini2018Data,idreos2013Big}. Examples include methods that enable efficient identification of the data subset of interest~\cite{lissandrini2018multi} and fast query execution over raw data sets through online aggregation~\cite{hellerstein1997online,Rusu-OLARAW}, result reuse~\cite{Galakatos:ResultReuse}, or dynamic prefetching~\cite{battle2016dynamic}. All these techniques aim to quickly provide query results to users to enable efficient selection of interesting data sets. Furthermore, there exist approaches that recommend interesting data sets to the user based on their previous information needs~\cite{vartak2015s,chirigati2016data}. Another key operation in this stage is the selection of relevant attributes of the data, often referred to as feature selection~\cite{tang2014feature}. This phase is of critical importance when the number of attributes in a data set is high and poses challenges for any downstream analysis. During such operations, analysts evaluate the value and importance of features both through their domain knowledge and through the use of metrics, e.g., variance, entropy, as heuristics.

\noindent{\textit{\textbf{Opportunities.}}}
For progressive data science, it will be interesting to extend data selection in the direction of active techniques that trigger downstream operations, such as data cleaning, if new relevant data becomes available. This is similar to the notion of publish-subscribe systems, where users subscribe for certain interesting data items and get notified actively once relevant data is becoming available. For example, in the medical domain, a doctor can register to be informed if entries for patients with a certain disease are being added to the database. Moreover, for progressive data science, downstream operations such as data preprocessing (Section~\ref{sec:kdd_preprocess}) and model re-training (Section~\ref{sec:kdd_datamining}) can be actively triggered based on such events. In particular, when feature selection is performed progressively, one big potential in downstream analysis is the ability to vary feature selection and build several models concurrently (consider the parallel lines in Figure~\ref{fig:kddPipeline}). Since the utility of these models can be approximated gradually in shorter time, it essentially allows analysts to investigate multiple hypotheses having different degrees of confidence. Our \textit{customer segmentation scenario} is a great example where progressive data selection can enable an analyst to quickly investigate combinations of multiple data sets for the clustering of users. For instance, an analysts can identify two large disjoint data sets, populated by two different operations in their organization, e.g., marketing and sales, perform a join of these data in a progressive manner, run the clustering algorithm and assess whether these data sources need to be fully integrated depending on the preliminary results from the progressive clustering computations.

\noindent{\textit{\textbf{Implications.}}}
Since data selection is the first stage in the KDD pipeline, the implications of a progressive approach reverberate throughout the entire process. On one hand, this is beneficial because the process can start much earlier and insights become available gradually. This facilitates a better steered KDD process. On the other hand, the data scientists have to be made fully aware that the derived insights are based on partial data and should be treated accordingly. We argue that, in fact, this is how most statistically-driven science is nonetheless performed---results are derived from studies performed over carefully selected samples. If the sample turns out to be inaccurate, the entire downstream pipeline is corrupted.

\noindent{\textit{\textbf{Challenges.}}}
Two sets of challenges are triggered by progressive data selection. The first relates to \textit{data-level challenges} that involve identifying data subsets that maximize the added information value throughout the downstream pipeline stages. This has to be done iteratively and requires a feedback loop. Moreover, a common measure has to be defined across the heterogeneous stages of the pipeline. The second set relates to \textit{efficiency challenges}. With progressive execution, the pipeline has to be evaluated multiple times. Complete reevaluation of the entire pipeline is likely inefficient. Thus, incremental execution strategies that integrate partial results have to be explored. While such solutions exist for homogeneous environments, e.g., relational databases, it is still a challenge how to extend them to the KDD pipeline.

\subsection{Data Preprocessing}
\label{sec:kdd_preprocess}
Data preprocessing in the KDD pipeline aims to identify and address quality issues in the data that are selected as interesting. Operations such as the identification of missing values and their imputation, removal of duplicate or problematic records, as well as the identification of outliers are typical for this stage~\cite{kandel2011research}. This is one of those stages where a significant amount of time is spent due to inconsistencies in the way data is gathered or stored.

\noindent{\textit{\textbf{Opportunities.}}}
When conducted in a progressive manner, where, for instance, new data is being made available continuously, some of the key data quality notions might deviate significantly. For instance, with new data being available, new missing value characteristics might emerge, or the scripts that are written to identify and fix data quality issues (e.g., for parsing certain numeric values and for converting them into a unified form) could fail with a dynamically changing representation of such values. The challenges are amplified when models of data are used to fix some of the data quality issues~\cite{collins2001comparison}. For instance, where missing values are replaced with the sample average of a feature, or where outliers are flagged based on the distribution characteristics (e.g., those that fall outside the 1.5 times the inter-quartile range), the varying characteristics of the data over the progressive process pose challenges. To illustrate, in our \textit{customer segmentation scenario}, the analyst can consider leaving out the ``super customers'' whose spending volume is considerably higher compared to the rest. This can be done by identifying outliers using a simple statistical thresholding approach where customers spending more than $x$ times the standard deviation of the distribution of per-customer spending are labelled. A progressive approach here would enable an analyst to investigate a range of $x$ values and observe how dramatically the resulting set is changing, eventually leading towards a more informed, effective decision on how to conduct this data processing step.

\noindent{\textit{\textbf{Implications.}}}
Decisions made at this stage often have significant implications for the stages that follow. In particular, in cases where new ``sanitised'' data instances are introduced into the data or where problematic records, e.g., outliers, are removed following the process discussed above, any further operation relies on the robustness of these decisions. Erroneous decisions made at this stage could easily bias and skew the models built on the data. One very common example is with missing value imputation and its impact on the data variance~\cite{troyanskaya2001missing}. Certain methods can amplify or reduce the co-variation between the data attributes and result in models that pick up on these artificial relations, such as a linear regression model getting stronger if the missing values are imputed following a linear model.

Progressive methods offer effective decision making when applying such critical operations on the data. Where there are several competing strategies to fix data quality issues, analysts can observe the downstream impact of these alternatives and choose those that introduce the least amount of bias. Being able to progressively observe and compare the consequences of a data-level operation on a further modelling stage leads not only to more efficient, but also better-informed decision-making in this stage.

\noindent{\textit{\textbf{Challenges.}}} 
There are some concrete challenges at this stage for progressiveness. \textit{Algorithmic challenges} stem from the fact that not all preprocessing operations are suitable for progressive computations. For instance, running even outlier analysis on progressive settings requires adaptive algorithms that align the criteria to the characteristics and the coverage of the parts of the data being considered. Similarly, consider the missing value imputation process in the progressive setting, which requires a novel way of thinking to adapt to the changing nature of missing values as new data becomes available, e.g., what is considered to be a missing at random at first could be deemed more systematic as more data becomes available. \textit{Communication challenges} occur in this stage with the need to effectively and transparently communicate the changes to data. With the fast paced interaction that progressiveness facilitates, effective communication of how the data is changing based on the interactive processing of the data is of key importance to enable analysts to explore several transformations in fast cycles.

\subsection{Data Transformation}
\label{sec:kdd_transformation}
In the transformation step of the KDD pipeline, the preprocessed data is modified and reorganized. Ideally, the transformed data becomes better suited as input to the data mining technique that follows, e.g., by removing redundant features or by deriving new ones.

\noindent{\textit{\textbf{Opportunities.}}}
Data transformation techniques heavily depend on the form in which the data is provided as input. If changes are applied in previous steps of the pipeline, the transformation generally holds as long as the number of features and their types are not changed. In a classification setting, for example, if mislabeled data are removed in the preprocessing step, the computations performed here remain unchanged. The scenario is drastically different if features are not removed. Consider, for example, in our customer segmentation scenario that each customer's address is recorded with GPS coordinates. As feature transformation, a function that maps coordinates to geographical entities such as regions, states, or nations is defined. If the GPS coordinates are dropped in the data selection phase, the transformation becomes ill-posed. Different strategies can then be adopted to deal with that scenario, such as stopping the progressive computations in the pipeline and informing the user. Another possibility is to ignore the computation of the derived feature and propagate only the ones not affected by the missing input.

\noindent{\textit{\textbf{Implications.}}}
Changes in the transformations applied to the data deeply affect the computations performed in later stages of the pipeline, often requiring a change in the data mining algorithms used. Changing the size of the geographical entities in the previous example, e.g, by transitioning from regional to state aggregation, may drastically affect the performance of the data mining or machine learning techniques that follow. 

The careful combination of the transformation function and the data mining technique is a cornerstone of the progressive data science pipeline. By directly reflecting the changes applied to the transformation functions, the user can fine-tune model performance by providing better conditioned data.

\noindent{\textit{\textbf{Challenges.}}}
An interesting \textit{procedural challenge} is worth noting here that relates to how the progressiveness is executed. This challenge occurs when the decision about the suitability of a particular transformation can be made only after an evaluation of all the downstream impacts of that operation, e.g., cases where you can only assess the suitability of a transformation after attempting to run a model with the data and observe accuracy. To alleviate this, intelligent ``data sensing'' methods could provide interactive feedback, e.g., progressively checking if transformed data adheres to particular statistical characteristics or if the underlying variation in the data is preserved robustly following the data reduction operations.

\subsection{Data Mining}
\label{sec:kdd_datamining}

The data mining step aims for the inference of a model from a given data set, which can help answer specific questions of the user, such as cluster formation, frequent pattern, or outlier detection. Depending on the specific task, different algorithms and models are used, e.g., deep supervised neural networks are suitable for image classification according to given categories, generative adversarial networks allow the generation of new realistic images. Unsupervised models such as clustering or visualization techniques enable an intuitive inspection of the data. The data mining step itself typically summarizes three sub-tasks: selection of an algorithm, underlying form of the model and objective function; selection of model and training meta-parameters; and parameter optimization based on the given data (training). While a large variety of algorithms for training exist, in particular the first step, selection of an objective and a principled form of the desired model, often depends on the intuition of the modeller: it constitutes the critical step to transform the often yet unclear objectives of the practical application into an exact mathematical objective which can be be solved  algorithmically. 

\noindent{\textit{\textbf{Opportunities.}}}
Progressive technologies can address several challenges in this context: foremost, model inference is often a time consuming process. Progressive techniques can help make advanced data analysis methods accessible in interactive settings where the mere computational complexity renders its classical form infeasible. This is the case for deep learning models, for example, but also support vector machines~\cite{cortes1995support} face a possibly quadratic kernel matrix, or random forests might be based on  a large number of components ~\cite{breiman2001random}.
Further, progressive modeling carries the potential to interactively shape parts of the data mining pipeline, which are not easy to formalize: this is the case for  multi-criteria settings, that means besides a primary objective such as the classification accuracy, characteristics such as model complexity or interpretability play an important role. In this case, a suitable compromise how to weight the parts needs to be found - an endeavor, which is often simpler if there exists the possibility of an efficient interactive trial and error. 

\noindent{\textit{\textbf{Implications.}}}
A change in the input data can affect all the sub-steps of data mining process, and it can do so in unprecedented ways with regard to computational complexity and accuracy of the results: one challenge is raised by an increase in data set size; this often requires an adaptation of model parameters, to account for the additional information. This is easily possible for local  models such as a kNN (k-nearest-neighbor) classifier, since the effect of single points is explicit in such models. The problem is not so easy for distributed representations such as deep networks, where single examples can affect the outcome in unprecedented ways.

In addition to model parameters, model meta-parameters such as the model complexity or number of clusters can also be affected. Notice that classical results from statistical learning theory often guarantee a consistency of the models with increasing data set size, i.e., model adaptation becomes less severe the more data is integrated; yet, these guarantees rely on the often unrealistic assumption of data being independently and identically distributed (i.i.d).

Another challenge occurs if the data representation changes because features are added to or deleted from the data. Feature-centered models, such as decision trees, allow the integration of additional features easily. Alternatives, e.g., deep networks require retraining, a usually time-consuming process. In all cases, major changes of the model meta-parameters, model architecture, and learning pipeline can occur, and a novel evaluation and interpretation of the results in the subsequent steps of the KDD pipeline becomes necessary.

\noindent{\textit{\textbf{Challenges.}}}
Concrete challenges, which need to be addressed in this context, include different aspects: 
\emph{Algorithmic challenges} occur, since model adaptation does no longer happen in batch mode, but incrementally, i.e.\ an existing model needs to be adapted according to novel data
\cite{DBLP:journals/ijon/LosingHW18}. The crucial question arises how this can be done efficiently without retraining the complete model and storing all training data.

\emph{Modelling challenges} face the question how to determine crucial model meta-parameters without having access to the complete data. The answer is simple, if restricted models such as linear models are used, or data-adaptive non-parametric models such as a kNN are used. Yet, in general, the question cannot be answered prior to learning, since an unexpected data complexity might arise over the course of time. Essentially, this means that model meta-parameters such as the degree of regularization or complexity of the model become model parameters, and efficient
ways how to deal with the stability-plasticity dilemma of machine learning need to be determined.

\emph{Usability challenges} arise due to the fact that the user deals with a model, which changes in its functionality and degree of validity, and the way it does so is possibly not expected. Thus, in addition to the mere model functionality, design has to account for human readability of learning, and it is advisable to e.g\ enforce a certain amount of stability and monotonicity of model behavior, to limit unexpected changes.

\subsection{Interpretation / Evaluation}
\label{sec:kdd_evaluation}
The interpretation/evaluation step aims to determine a measure of quality or performance of the data mining model. The goal is to assess the observations, patterns and modeling results in terms of their value and validity. This can incorporate external information, such as class labels, human assessment of the model results, or the use of internal information to the model, such as the objective function it optimizes.

\noindent{\textit{\textbf{Opportunities.}}} 
When all previous steps are done in a progressive way, the progressive data science pipeline should support human intervention for two different kinds of time-varying information:
\begin{itemize}
    \item Progression of data mining results, and
    \item Progression of model evaluation measures
\end{itemize}

The former provides direct information of the data mining output, while the latter represents meta information of the process evolution. 
Regarding the evolution of model evaluation measures, the user can make the decision for an early termination once a monotonic curve is produced, indicating the model with low chances of getting better performance. Moreover, changes in the early steps of the pipeline, e.g., in the data transformation step, may also be evaluated here. For example, fine tuning of the feature space may lead to better optimization of the objective function.

\noindent{\textit{\textbf{Implications.}}}
Being a stage that is often considered towards the end of an analysis session, observations made at this stage has often upstream implications, i.e., earlier stages in the analysis process could be questioned due to the poor quality of the results. Alternatively, the analyst might decide to initiate ``parallel'' modelling processes where certain decisions are varied along the pipeline and the result are compared. For instance, in our \textit{customer segmentation scenario}, analyst might decide to vary some parameter of the algorithm (such as the $k$ in the \textit{k-means} algorithm~\cite{macqueen1967some}) and compare the quality or validity of the results in these two different ``progressive runs'' of the method. Some observations made at this stage may even lead to changes all the way to the beginning of the process, for instance, due to a lack of improvement in the progressive results, the analyst might decide to restart the whole pipeline by considering a new data set that was initially left out from the analysis, leading to a complete overhaul of the computations.

\noindent{\textit{\textbf{Challenges.}}} 
A key \textit{decision-making challenge} arises here for supporting analysts in making informed judgments on the quality of the results. In order to effectively evaluate a result where indicators of quality are approximate in a progressive setting, effective heuristics that quantify the uncertainty in the results, and the level of convergence (towards a final result) are needed to be estimated and communicated. Potential ad hoc heuristics could be the rate of change in the overall model over iterations~\cite{turkay2017designing}, or percentage of data processed~\cite{schulz2016enhanced}. However, further research is needed to develop generalizable, systematically evaluated heuristics for uncertainty and convergence of algorithms to serve as effective evaluation criteria in progressive settings.

%The data mining results are intrinsically difficult to analyze, often being high-dimensional data that cannot be directly visualized on screen. Therefore, the progressive pipeline must combine visualization techniques that are able to deal with time varying, uncertain, and high-dimensional data. While this is a challenging and open problem, we think that is an important step for the effective implementation of the data science pipeline.

\subsection{Putting it all together}
\label{sec:kdd_all}
As discussed in all the stages above, progressiveness brings new opportunities when the widely adopted KDD pipeline is reconsidered through this novel lens. In many of the cases above, progressiveness facilitates a fast-paced, flexible, and adaptable analysis process that empowers analysts in dealing with large, heterogeneous, and dynamic data sources, and in generating and evaluating hypotheses and insights. We argue that a primary mechanism that enables this analysis paradigm is the ability to quickly propagate the results of any stage to downstream steps, observe the resulting impact as early as possible, and make changes to the early stage conditions to iterate further. With this very strength comes also the core challenge of progressive approaches---the inherent uncertainty introduced into the pipeline by progressive methods and how this uncertainty can be recognized and considered (see further discussions in Section~\ref{sec:discussion}). Suitable methods are needed to manage the progressive steps in ways where uncertainty at each stage is clearly decoupled and made transparent. Analysts also need methods where they can control and debug the whole pipeline in a seamless manner where they can iterate between the various stages fluidly both upstream and downstream.

%% file: Sec3-Promisingfirststeps.tex
\section{Promising First Steps}
\label{sec:firststeps}
In the following, we discuss existing approaches that are related to our vision from three different communities: database, machine learning, and visualization. As is demonstrated in this section, there is already a huge body of fundamental work existing in these different communities that can be leveraged to enable our vision of progressive data science. However, what is missing is a more holistic view that discusses new progressive approaches that cut through the individual steps of a KDD process and connect those steps to enable data scientists to revisit decisions in all steps and immediately see the effect of changes to all the other steps. One important long-term challenge is, thus, to bring all the existing individual results together in more open and connectable progressive systems that span over the complete KDD process and help data scientists to solve their problems more efficiently.

\subsection{Highlights from DB community}
\label{sec:db_highlights}
The database community has recently been working on aspects to make the individual steps of a KDD process more interactive. One major line of work is centered around query processing and tackles the question of how to enable database engines to provide interactive response times on large data sets. Motivated by \textit{data-level}, \textit{efficiency} and \textit{algorithmic} challenges outlined in Sections~\ref{sec:kdd_selection} and~\ref{sec:kdd_preprocess}, this line of work not only includes approximate query processing techniques~\cite{AQPBook} that use sampling to achieve interactivity, but also other query processing techniques that aim to re-use previously computed results in a user session (where database queries are potentially built incrementally)~\cite{Galakatos:ResultReuse,wasay2017data, dursun2017revisiting,DBLP:conf/edbt/GogolouTPB19}.

Another line of research has studied the problem of adaptivity, where the system adapts itself (e.g., the data organization, or the index structures), in order to execute queries in an efficient manner~\cite{idreos2012adaptive,zoumpatianos2016ads}. Furthermore, there also exist more advanced speculative query processing techniques~\cite{battle2016dynamic,DBLP:conf/icde/KamatJTN14}, which predict what the user is likely to look at next in order to start the computation eagerly. All the before-mentioned interactive query processing techniques are basic approaches that can help to speed up different KDD steps. For example, sampling-based query processing is not only used in the initial data exploration step to help users identify relevant data faster \cite{DBLP:conf/sigmod/CrottyGZBK16}, but also for making data mining and model building approaches more efficient~\cite{Rusu:SGDOLA,DBLP:journals/pvldb/Kraska18,DBLP:journals/pvldb/SunRYD14}.
Moreover, there also exist other lines of research in databases not centered around query processing that can be used to make other KDD steps more interactive. One important line is on interactive data cleaning and wrangling~\cite{kandel2011wrangler,DBLP:conf/sigmod/WangKFGKM14,kandel2011wrangler,le2014flashextract, jin2017foofah} to support more efficient extraction of structured data from semi-structured data sets building on the opportunities outlined in Section~\ref{sec:kdd_selection}. 

Another line of work that is important is on recommendation algorithms for data exploration that suggest potentially interesting insights, enabling an easier understanding of large and new data sets~\cite{vartak2015s, chirigati2016data}. Furthermore, there exist many directions on related areas such as benchmarking interactive database systems \cite{DBLP:conf/sigmod/BattleABCEFSSW18,DBLP:journals/debu/EichmannZZBK16}, but also on making data exploration more safe and avoid that data scientists ``tap'' into typical statistical pitfalls~\cite{Binnig2017toward,DBLP:conf/sigmod/GuoBK17} as also discussed as an opportunity in Section~\ref{sec:kdd_transformation}. An interesting fact that manifests that interactivity and progressiveness play an important role in the database community is the fact that there are multiple workshops co-located with major conferences (e.g., HILDA @ SIGMOD\footnote{http://hilda.io/}, ExploreDB @ SIGMOD\footnote{https://sites.google.com/a/unitn.it/exploredb18/}, and IDEA @ KDD\footnote{http://poloclub.gatech.edu/idea2018/}). All these workshops foster new results on the problems related to the above mentioned areas.

\subsection{Highlights from ML community}
\label{sec:ml_highlights}
Humans' extraordinary mental plasticity enables the seamless life-long learning and efficient incremental adaptation of natural intelligence to novel, non-stationary environments. Yet, one of the major challenges of artificial intelligence remains the question how to efficiently leverage learned strategies to novel environments. Albeit this question is widely unsolved, quite a few promising approaches exist, which carry a high potential as major ingredients of progressive data analytics. Motivated by the \textit{algorithmic challenges} outlined in Section~\ref{sec:kdd_datamining}, \emph{incremental and life-long learning} architectures, as an example, address the question how to efficiently adapt data mining models such that they become consistent to novel data, even if the latter might be subject to concept drift, that means the underlying data distribution changes significantly in comparison to previous observations~\cite{gama2014survey}. Interestingly, it is possible to set up methods which can efficiently and agnostically deal with a large variety of different types of drift, by using either active drift detection, robust ensemble methods, or intelligent memory structures to efficiently face the stability-plasticity dilemma~\cite{losing2016knn}. These machine learning technologies can serve as key ingredients whenever the size of the data set changes in progressive data analysis, with open source tool-boxes for such streaming data analysis being readily available, such as the  MOA framework~\footnote{https://moa.cms.waikato.ac.nz/} by Bifet et al.~\cite{bifet2010MOA}. These technologies from the ML community have a direct impact on the dedicated Data Mining step of the KDD pipeline (building on the opportunities as outlined in Section~\ref{sec:kdd_datamining}), since they enable us to exchange data mining modules, which stem from ML, by incremental approaches. 

Beyond this direct impact, ML approaches can also help facilitating data pre-processing and transformation (step 2 and 3 of the KDD pipeline) in progressive settings, by offering technologies, which can cope with changes in the data representation.
As an example,
the question on how to deal with changing data representations or tasks is addressed in so-called \emph{transfer learning}~\cite{pan2010survey}: how can an existing model be transferred to either a different task or a different data representation, thereby preserving relevant common structural principles? In this realm, quite a few promising technologies offer interesting ingredients for progressive data analytics pipelines. This includes fast adaptation technologies to transfer a model to a novel probability density function~\cite{courty2017optimal} and progressive neural networks for efficiently learning strategies in reinforcement settings~\cite{rusu2016progressive}. A third example are \emph{representation learning} technologies which aim for invariant data representations which enable its seamless use for a wide range of different settings~\cite{evangelopoulos2014learning}, whereby universal representations as offered, e.g., by deep networks tool-kits, are freely available for important domains such as vision~\cite{zacharias2018survey}.

The increasing availability of big, often unlabeled data sets as well  possibilities to learn extremely realistic generative models e.g.\ based on generative adversarial networks has generated a boost of the area of \emph{active learning} in ML pipelines
\cite{DBLP:journals/corr/ZhuB17}. Essentially, these approaches propose which samples to include into a modelling framework based on the objective to maximally increase the accuracy and confidence of a learned model. These approaches offer promising support for the first step of the KDD pipeline, since they highlight, which information might provide the maximum gain for the formal models. 

Interestingly, there is a clear mutual benefit by integrating aspects of progressiveness in ML and data analytics, since this liaison adds the novel crucial challenge of human-readability and usability to the primarily algorithmic challenges faced by ML technologies. There do exist  approaches how to increase readability in physical interaction scenarios \cite{DBLP:journals/arobots/HuangHAD19}, and \emph{human readability of ML models} is extensively discussed, yet not fully solved, in the realm of \emph{explainable AI} \cite{DBLP:journals/corr/abs-1806-00069}. We expect that progressive data analytics, in particular interactive ML \cite{DBLP:journals/corr/LeeJC16} will have a significant impact to further this important research domain.

\subsection{Highlights from VIS community}
\label{sec:visualisation_highlights}
Building progressive visualization and visual analytics systems for data science currently requires complex and expensive developments since existing systems are not designed to be progressive. There has been a few prototypes of progressive visualization and visual analytics systems that proved that the approach was useful and effective for analysts, but they currently remain ad-hoc and monolithic. We will review the most popular ones. Recently, there has been some attempts at building infrastructures natively progressive from the ground up. Although the work is still ongoing, more work will be needed to design and implement fully progressive systems at the level of eager ones such as R or Python with their data science stack. While the work is only starting~\cite{fekete2016progressive, crotty2015vizdom}, it offers a huge potential for research and opportunities for building scalable interactive data science systems where the whole pipeline is addressed as discussed in Section~\ref{sec:kdd_all}.

The main idea that progressive systems can help human carry long-lasting cognitive tasks has been validated by a study by Zgraggen et al.~\cite{zgraggen2017progressive} showing that while human attention is hurt by latencies over 10 seconds, providing progressive results every 1-5 seconds instead of instantaneous results allow analysts to perform exploratory tasks with a similar level of attention. Another experiment by Badam et al.~\cite{badam2017steering} confirmed that analysts can perform complex analyses using a progressive system, understand when to make decisions or when to refrain from making decisions, and interact in a complex way with a progressive system while it is running. However, the experiment was performed using a prototype system where the results of the algorithms were pre-computed to control their latency. These investigations are some early signs that the \textit{usability} and \textit{decision-making} challenges outlined in Sections~\ref{sec:kdd_datamining} and~\ref{sec:kdd_evaluation} are being empirically observed and addressed in ongoing visualization research.

Progressive visualization systems are popular for graph visualization where many graph layout algorithms are iterative. Systems like Tulip~\cite{auber2017tulip} and libraries likes D3~\cite{bostock2011d3} implement progressive graph layouts that are popular and allow moving nodes while the algorithm is running to steer the layout. MDSteer~\cite{williams2004steerable} provide a similar system for multidimensional scaling, also allowing users to focus on areas of the visualization to steer the computation. However, until recently, progressive visualization was limited to iterative layout algorithms. More recently, several visual analytics applications have been built to deliver progressive results in efforts to address communication, usability and decision-making challenges as discussed in Section~\ref{sec:challenges}. Stolper et al.~\cite{stolper2014progressive} coined the term``Progressive Visual Analytic'' (PVA), presented the paradigm, explained through an example application ``Progressive Insight'' meant to mine event sequences to find interesting patterns. The publication has been followed by studies and requirements for PVA~\cite{muhlbacher2014opening, badam2017steering}, by systems performing various kinds of progressive analyses~\cite{badam2017steering, hollt2016cytosplore}, by techniques facilitating progressive computational modelling~\cite{turkay2017designing}, and by articles describing progressive ML algorithms, such as t-SNE~\cite{pezzotti2016hierarchical, pezzotti2017approximated}, k-nearest neighbors, regression, density estimation~\cite{jo2018panene}, and event sequence pattern mining algorithms~\cite{stolper2014progressive, servan2018prosecco, raveneau2018Progressive}.

%% file: Sec4-Discussion.tex
\section{Discussion}
\label{sec:discussion}
While we present progressiveness as a promising approach that can transform the way that data analysis is conducted, the approach is not without its limitations. As with any algorithm that is trying to infer from partial data, progressiveness by nature provides approximate, hence, uncertain results~\cite{Joseph1999, Fisher2012} that are likely to contain errors~\cite{Ding2016}. Analysts working in progressive settings thus need to recognize and work effectively with the uncertainty along the pipeline. Research on progressive data science need to consider the uncertainty challenge carefully, and investigate ways to both minimize and/or control it and develop methods for the effective communication of uncertainty at each stage of the pipeline. User specified error-bounds to balance utility and accuracy~\cite{Ding2016}, data sampling techniques that preserve specific characteristics such as the ordering of categories~\cite{kim2015rapid} in visualizations, or prioritizing user defined ``interesting''~\cite{Raman1999} data instances and important features~\cite{Rahman2017} are good examples of controlling the error in the process effectively. When it comes to the communication of uncertainty, error bars with confidence intervals have been used commonly~\cite{Joseph1999, Fisher2012}, although they have shown to be confusing for some analysts~\cite{Fisher2012}. There are recent activities within visualization research that aims to understand better how people interpret uncertain information~\cite{Fernandes2018} and how can the uncertainty in the statistics be better explained through alternative representations~\cite{correll2014harmfulbars}.  Progressive data science solutions need to incorporate and advance such techniques to support analysts in making analytical decisions under uncertainty at ``all'' the stages of the pipeline to provide comprehensive support and guidance.

In addition to this challenge on communication of uncertainty, the quantification of the quality of the progressive results is also an important challenge. The estimations of progression quality might not always be accurate or hard to estimate, in particular with certain algorithms that are by design not suitable for progressive computations. The quantification/estimation of progression and its quality is one challenge that requires further research.

Similarly, not all tasks are equally effectively addressed by progressive approaches and there are differences how well different algorithms perform as also highlighted in the survey by~\cite{DBLP:journals/ijon/LosingHW18}. This is due to the expectations from an effective incremental learning approach that the model has to adapt to changes in data gradually with having access to only a partial set of the data with the capability to return approximate results \textit{anytime}~\cite{DBLP:journals/ijon/LosingHW18}. For tasks (such as finding MIN/MAX of a data column~\cite{li2018approximate}) where approximate answers are not desired or analysts are not able to effectively make decisions, progressive methods might not be the most suitable. One approach here could be to investigate ways to conduct the certain phases of the data science pipeline in progressive ways and only resorting to batch computation when needed. For instance, analysts can perform the data-related, and model tuning tasks progressively and by accepting a certain degree of error, but only then resort to more accurate, batch computation for the final model training phase. Designing for progressiveness is an area that is ripe for research and developing progressive counterparts for traditionally non-iterative, batch based algorithms~\cite{jo2018panene} offers opportunities for further research. 

With progressiveness, we naturally argue for an analysis approach where the analyst has a much more active, involved and critical role in the data science process. As opposed to trends that try to minimize human involvement in data science~\cite{Feurer2015, Kanter2015}, the progressive data science approach considers the involvement of the human as a strength. By empowering the analyst in the process, we think that the resulting models are better informed by the domain expertise, much more understood and trusted by those who are building them. This resonates well with visual analytics research~\cite{Keim2008}, however, as discussed in Section~\ref{sec:visualisation_highlights}, involvement of the user in progressive settings through interactive interfaces is still in its early days. We call for further research on areas such as uncertainty communication, communication of progression, specialized interaction methods and metaphors to capture user input more effectively.

One particular challenge that stems from using multiple progressive approaches within a data science pipeline is related to the management of parallel computation streams~\cite{turkay2017designing}, e.g., computations running in parallel that have different convergence rates, and keeping a record of all the different operations made in the progress, which could be broadly referred to as analytic provenance~\cite{ragan2015characterizing}. Effective methods are needed to have an oversight of all the alternative routes investigated by the analyst, and research in the visualization of the sensemaking process~\cite{xu2015analytic} has potential applications here. However, most of the research so far has focused on batch processing settings and novel methods that can handle progressive computations need to be developed. 

%% file: Sec5-Conclusion.tex
\section{Conclusion}

With this paper we introduce \textit{Progressive Data Science} as a new paradigm where analysts are provided with approximate yet informative, intermediate responses from computational mechanisms in short time anywhere within the analysis pipeline. 
In this approach, letting the analysts interact with the intermediate results allow an early detection of wrong or suboptimal choices, and offer significant improvements within the iterative, traditionally trial-and-error based stages of data science process. In this paper, we presented a unifying vision through a rethinking of the widely adopted and influential KDD pipeline and discussed the various challenges arising from progressiveness followed with a discussion on the promising first steps from different communities where progressive methods are of interest.

We propose \textit{Progressive Data Science} as a novel knowledge discovery paradigm where progressiveness is inherent in every step of the data science process, and ensuring success in such a novel paradigm requires the concerted effort from various research communities. There are several already promising first steps from different communities that demonstrate the potential of the approach, and many interesting scientific challenges lie ahead which require multidisciplinary thinking. We are confident that teams of researchers from complementary domains will address these challenges to further establish this paradigm. 